\documentstyle[aps,pre,preprint]{revtex}
\tightenlines
\begin{document}
\author{M.S. Hussein and M.P. Pato}
\address{Nuclear Theory and Elementary Particle Phenomenology Group\\
Departamento de F\'{\i}sica Nuclear, Instituto de F\'{\i}sica, Universidade de
S\~{a}o Paulo\\
CP 66318, 05315-970 S\~{a}o Paulo, SP, Brazil}
\title{Distribution of Matrix Elements of Random Operators\footnote{Supported 
in part by the CNPq and FAPESP (Brazil)}}
\maketitle

\begin{abstract}
It is shown that an operator can be defined in the abstract space of a
random matrix ensemble whose matrix elements statistical distribution
simulates the behavior of the distribution found in real physical systems.
It is found that the key quantity that determine these distributions is
commutator of the operator with the Hamiltonian. Application to symmetry
breaking in quantum many-body system is discussed.
\end{abstract}

\newpage 

Ensembles of random matrix theory (RMT)\cite{Meht,Brody,Weide} have had a
wide application as models to describe statistical properties of eigenvalues
and eigenfunctions of chaotic many-body systems. More general ensembles have
also been considered\cite{Dyson}, in order to cover situations that depart
from the conditions of applicability of RMT. One such class of ensembles is
the so-called deformed Gaussian orthogonal ensemble (DGOE)\cite{Pato1}.
These ''intermediate '' ensembles are particularly useful when one wants to
study the breaking of a given symmetry in a many-body system such as the
atomic nucleus. Further, different variances of the deformed random matrix
ensembles can interpolate any one of the three universal RMT ensembles,
i.e., the GOE, the GUE and the GSE, and the Poisson ensemble, which would
represent the transition between a fully chaotic situation, with no
conserved quantity but energy, to a regular one, i.e., one with a complete
set of operators that commute with the Hamiltonian. When discrete symmetries
such as isospin, parity and time reversal are violated in a complex
many-body environment,one relies on a description based on the transition
between one GOE\cite{Pato2,Guhr} into two coupled GOE 's in the first two
cases and a GOE into GUE in the last case. This latter case has recently
been studied by us in the case of disordered metal-insulator transition\cite
{Patoh}.

The above mentioned statistical properties refer to fluctuations, around
average values, of quantities connected to the eigenvalues and the
eigenfunctions. These mean values are specific to the physical system being
considered and, in fact, semiclassical estimates of them can be derived in
terms of the underlying Hamiltonian\cite{Peres}. The fluctuations, on the
other hand, have a quantum origin and are, in principle, universal, in the
sense that the only information they carry about the system, is the class of
underlying symmetry they belong. In order to compare the statistics
generated by these fluctuations with the predictions of the ensembles, it is
therefore necessary, in both cases, of eigenvalues and eigenfunctions, to
perform some convenient rescaling of variables that eliminates their average
behavior. In the case of the eigenvalues, they are first unfolded, which
means that they are mapped onto new levels with a constant density made
equal to one. For the eigenfunctions, the most convenient quantities to be
statistically analyzed are not directly the components of the
eigenfunctions, taken with respect to some particular basis, but rather
matrix elements of a given operator. These matrix elements also show
statistical fluctuations around their mean values\cite{Peres} and they have
to be subjected to some local average process that extracts secular
variations as a function of the energy\cite{Alha}. However, what is still
lacking is a direct comparison of matrix elements distributions with
ensemble calculations. The difficulty being that, except in the limiting
case of fully chaotic regime, nothing has been so far done in random matrix
ensembles studies about these distributions. In the chaotic situation, with
no conserved quantities and, as consequence, without any active selection
rule, the matrix elements behave like components of a isotropic random
vector and one expects them to follow the same distribution of the
eigenstate components, i.e., the Porter-Thomas law\cite{Porter}.
Nevertheless, this argument does not hold in intermediate situations between
chaos and order.

Of course, the reason why matrix elements distributions have not yet been
investigated, in the context of matrix ensembles, is simple: it is not clear
how one can define an operator associated to an observable in this abstract
space. This is exactly the question that is being addressed in this letter.
We want to show that it is possible to introduce some random operator whose
matrix elements would simulate, in some way, the behavior of observables
found in calculations and measurements performed in real physical systems.
In the construction of this operator, we will be guided by the idea that
when a system undergoes a chaos-order transition, a quantity that has a key
role in determining the statistical behavior of the matrix elements of an
operator, is the value of its commutator with the Hamiltonian.This is
implied by the equation

\begin{equation}
\left( E_{l}-E_{k}\right) <E_{l}\mid O\mid E_{k}>=<E_{k}\mid [O,H]\mid
E_{k}>=i\hbar \frac{d}{dt}<E_{l}\mid O\mid E_{k}>  \label{eq 1}
\end{equation}
here the last equality was obtained using Schr\"{o}dinger equation and
assuming an operator with no explicit dependence on time. Eq.$\left( \ref{eq
1}\right) $ clearly shows that the commutator supplies the connection
between the matrix elements of an observable and its behavior as a function
of time.

To see how this is reflected in the statistical distribution, suppose that
we choose to look at the matrix elements of an observable $O$ which becomes
a conserved quantity in the regular regime. The distribution of these
elements will undergo a transition from the Porter-Thomas law, at the
chaotic side, to the singular distribution $<E_{l}\mid O\mid E_{k}>$ $%
\propto \delta _{kl},$ at the regular side, since the last term in $%
\left( \ref{eq 1}\right) $ is zero in this case. In a more general
situation, as it occurs when reduced transition strengths are measured, what
happens is that some transitions might become forbidden at the less chaotic
regime. Then as the selection rules becomes operative, the first term in $%
\left( \ref{eq 1}\right) $ vanishes for some pairs $\left( l,k\right) $ and,
as a consequence, we may say that we have a partial conservation of the
observable which is causing the transitions. Again one should expect the
missing transitions to cause a deviation of the statistical distributions
from the Porter-Thomas law.

To define the ensembles of random matrices we are going to work with, we
follow the construction based on the Maximum Entropy Principle\cite{Pato1},
that leads to a random Hamiltonian which can be cast in the form

\begin{equation}
H=H_{0}+\lambda H_{1},  \label{eq 1b}
\end{equation}
where $\lambda $ is the parameter that controls the chaoticity of the
ensemble. We will assume it to be defined in the domain $0\leq \lambda \leq
1 $, in such a way that for $\lambda =1,$ $H=H^{GOE},$ and for $\lambda =0,$
we have some reduced chaotic situation defined by the choice of $H_{0}$.
Since we are specifically interested in the transitions from GOE to Poisson
and from GOE to two coupled GOE's, the above requirements are sufficient to
determine $H_{0}$ and $H_{1}.$

First, we consider the GOE$\rightarrow $Poisson transition, in which case we
write\cite{Pato0}

\begin{equation}
H_{0}=\sum_{i=0}^{N}P_{i}H^{GOE}P_{i}  \label{eq 2}
\end{equation}
and

\begin{equation}
H_{1}=\sum_{i\neq j}^{N}P_{i}H^{GOE}P_{j}  \label{eq 3}
\end{equation}
where $H^{GOE}$ is a $N$ - dimensional random matrix taken from the Gaussian
Orthogonal Ensemble, and we have introduced the projection operators $%
P_{i}=\mid i><i\mid ,$ $i=1,\ldots ,N$ . It is straightforward to verify,
from the usual properties of projectors, that $H=H^{GOE}$ for $\lambda =1$
and, on the other hand, when $\lambda =0$, $H$ becomes a diagonal matrix
whose eigenvalues are known to the follow Poisson distribution\cite{Meht}.

Considering now the GOE$\rightarrow $2GOE's transition, we write\cite
{Pato1,Pato2}

\begin{equation}
H_{0}=PH^{GOE}P+QH^{GOE}Q  \label{eq 4}
\end{equation}
and

\begin{equation}
H_{1}=PH^{GOE}Q+QH^{GOE}P  \label{eq 5}
\end{equation}
where $P=\sum\limits_{i=1}^{M}P_{i}$ and $Q=1-P.$ Here $H_{0}$ is a two
blocks diagonal matrix of dimensions $M$ and $N-M$ and each block is by
construction a GOE random matrix$.$ Again, it is easily verified that $%
H=H^{GOE}$ for $\lambda =1.$

Turning now to the operator, we choose it to have the form

\begin{equation}
O=\sum_{i=0}^{N}P_{i}H^{GOE}P_{i}.  \label{eq 6}
\end{equation}
In the case of the transition towards Poisson, since $O=H_{0}$, we
immediately derive the commutator relation

\[
\lbrack H,O]=\lambda [H_{1},O]. 
\]
where the commutator on the RHS has matrix elements given by

\begin{equation}
\lbrack H_{0},H_{1}]_{ij}=\left( H_{ii}^{GOE}-H_{jj}^{GOE}\right)
H_{ij}^{GOE}  \label{eq 8}
\end{equation}
which obviously is a nonvanishing antisymmetric random matrix. Therefore,
the statistical behavior of the elements of $O$ is controlled by $\lambda $
and it evolves from the Porter-Thomas distribution to a singular delta
distribution as $\lambda $ goes from $1$ to $0.$

On the other hand, for the transition to two GOE's, it is convenient to
separate the sum in Eq. $\left( \ref{eq 6}\right) $ into two parts in which
the first $M$ terms define the operator $O_{P}$ and the other $N-M$ define
the operator $O_{Q}$ , by construction $O=O_{P}+O_{Q}.$ The commutator with
the Hamiltonian can then be written as

\begin{equation}
\lbrack O_{P},PHP]+[O_{Q},QHQ]+\lambda \left(
O_{P}PHQ+O_{Q}QHP-QHPO_{P}-PHQO_{Q}\right)  \label{eq 10}
\end{equation}
These terms have a simple interpretation. The first and the second ones are
responsible for the transitions between states inside the blocks $PHP$ and $%
QHQ$, respectively. On the other hand, the four terms inside the parenthesis
cause transitions among states located inside different blocks. Since the
latter terms in (\ref{eq 10}) are all multiplied by the parameter $\lambda ,$
when $\lambda \rightarrow 0$ these transitions become forbidden. This
property shows that the operator we have introduced is very convenient to
study transition towards two coupled GOE's, a scenario appropriate to
investigate discrete symmetry violation in complex quantum systems.

To make our model more flexible we are going to consider, in this case of
the transition to two GOE's, matrix elements of the generic operator

\[
O^{\prime }=\left( 1-q\right) H_{0}+qO 
\]
where $H_{0}$ is given by Eq. (\ref{eq 4}) and $q$ varies between $0$ and $1$%
. With this form $O^{\prime }$ represents an operator which has a conserved
part. We have now a model with two parameters, the parameter $\lambda $ of
the Hamiltonian which may be fixed by the fitting the eigenvalues
distribution and the parameter $q$ that selects the operator. For $q=1$, we
have just a selection rule, as $q$ decreases we are introducing a
localization in the matrix elements inside the selection rule.

Following the standard procedure, we first construct, with the operator $O,$
the normalized vector

\[
\mid \alpha _{k}>=\frac{O\mid E_{k}>}{<E_{k}\mid O^{2}\mid E_{k}>} 
\]
where $\mid E_{k}>$ with $k=1,\ldots ,N$ is an eigenvector of the
Hamiltonian. From these $N$ vectors we define the matrix elements

\[
T_{kl}=<E_{l}\mid \alpha _{k}> 
\]
which are the quantities to be statistically analyzed.. It is convenient to
work with $\mid T_{kl}\mid ^{2},$ and, as mentioned above, perform a local
average that extracts secular variation with the energies. Thus we introduce
the quantities

\[
y_{kl}=\frac{\mid T_{kl}\mid ^{2}}{<\mid T_{kl}\mid ^{2}>} 
\]
where the average is done by using a Gaussian filter of variance equal to 2.
It has become standard in the analysis of these quantities to histogram
their logarithm. In the Figs. 2 and 3, it is shown the numerical results
obtained for the two transitions.

In Fig. 1, calculations performed in the case of the GOE-Poisson transition
are presented. The histograms of the logarithm of the matrix elements for
four different values of the chaoticity parameter $\lambda $ are plotted
together with three theoretical distributions: the Porter-Thomas,
corresponding to a $\chi ^{2}\left( \nu \right) $ distribution with $\nu =1$
degree of freedom , a $\chi ^{2}\left( \nu \right) $ where the degree of
freedom $\nu $ is derived directly from the ''data ''\cite{Alha} and finally
a distribution in which the histograms were fitted with the superposition of
two $\chi ^{2}$ distributions. Whereas one $\chi ^{2}$ distribution of the
type suggested in Ref. \cite{Alha}, is constrained to a peak always around
zero, our calculations, however, suggest the need of a linear combination of
two distribution as we have already proposed in Ref. \cite{Pato2}. This kind
of behavior has seems to be typical of transitions in which the eigenstates
become more and more localized. It can be understood as a signature of the
multifractal nature of the states\cite{Varga}.

In Fig. 2, the results for the transition towards two GOE's are shown. The
chaoticity parameter was fixed at value $\lambda =0.032$ and the parameter $%
q,$ that measures the localization inside the selection rule, is varied.
Again, the same theoretical distributions of Fig. 1 are also shown. At this
value $\lambda $, we expect to be near the case in which the two GOE's are
completely decoupled. We see from the figure that the distributions are
greatly dependent on the parameter $q$.

We observe that in the extreme situation when we have two diagonal uncoupled
blocks, the probability distribution can be written down explicitly. In
fact, since inside each block we have Porter-Thomas distributions and
outside of them the matrix elements vanish, we have for a generic reduced
strength $y=\mid T_{kl}\mid ^{2}$, the distribution

\[
P\left( y\right) =\left( \frac{M_{1}}{N}\right) ^{2}\sqrt{\frac{\pi M_{1}}{2y%
}}\exp \left( -\frac{M_{1}y}{2}\right) +\left( \frac{M_{2}}{N}\right) ^{2}%
\sqrt{\frac{\pi M_{2}}{2y}}\exp \left( -\frac{M_{2}y}{2}\right) +4\frac{%
M_{1}M_{2}}{N^{2}}\delta \left( y\right) 
\]
where $M_{1}+M_{2}=N$. The case $q=0$, in Fig.2, corresponds, for the above
value of the chaoticity parameter, to an almost uncoupled situation. We see
that we have only a slight deviation from Potter-Thomas. This means that the
delta function, in the above expression, is practically not observed. This
can explained by the fact that we are plotting the logarithm of the
intensities. In this kind of plot the zero elements are not observed and,
more than that, they gain a vanishing statistical weight that comes from the
Jacobian of the transformation $y\rightarrow \ln y$. Physically this means
that the selection rule is hardly detected in this kind of analysis if what
is being considered are statistics of matrix elements of an operator without
a conserving part.

In conclusion, we have extended in this paper the Maximum entropy theory to
the case of the distribution of matrix elements. Contrary to what have been
suggested recently in the literature, we find that in the intermediate
situation described by the deformed Gaussian ensembles, the distribution is
like a sum of two $\chi ^{2}$ ones. Our theory is quite fit to address the
question of symmetry breaking in complex many-body systems. The application
too isospin symmetry violation in light nuclei\cite{Mitch} is underway\cite
{Phmit}.

\ {\bf Figure Captions:}

Fig. 1 Four histograms of the logarithm of the matrix elements distributions
of the random operator O, see text, in the case of the transition GOE 
$\rightarrow$ Poisson, for the indicated values of the chaoticity parameter 
$\lambda.$ The calculations were done with matrices of dimension $N=100.$ 

Fig. 2 Four histograms of the logarithm of the matrix elements distributions
of the random operator $O^{\prime}$, see text, in the case of the transition 
GOE $\rightarrow$ 2GOE's, for the indicated values of the parameter $q.$ 
The calculations were done with matrices of dimension $N=120$ and block
sizes $M_{1}=40$ and  $M_{2}=80.$


\begin{references}
\bibitem{Meht}  M.L. Mehta, {\it Random Matrices} 2nd Edition (Academic
Press, Boston, 1991).

\bibitem{Brody}  T.A. Brody et al., Rev. Mod. Phys.{\bf \ 53}, 385 (1981).

\bibitem{Weide}  For a recent review, see T. Guhr, A. M\"{u}ller-Groeling
and A. Weidenm\"{u}ller, Phys. Rep. {\bf 299, }189 (1998).

\bibitem{Dyson}  F.J. Dyson, J. Math. Phys. {\bf 3}, 1191 (1962).

\bibitem{Pato1}  M. S. Hussein, and M.P. Pato, Phys. Rev. Lett. {\bf 70, }%
1089{\bf \ }(1993).

\bibitem{Pato2}  M. S. Hussein, and M.P. Pato, Phys. Rev. {\bf C 47,} 2401
(1993).

\bibitem{Guhr}  T. Guhr and, H.A. Weidenm\"{u}ller. Ann. Phys. (NY), {\bf 199%
}, 412 (1990).

\bibitem{Patoh}  M. S. Hussein, and M.P. Pato, Phys. Rev. Lett. {\bf 80, }%
1003{\bf \ }(1998).

\bibitem{Peres}  M. Feingold, N. Moiseyev and A. Peres, Chem. Phys. Lett. 
{\bf 117}, 344 (1985).

\bibitem{Alha}  Y. Alhassid, and R.D. Levine, Phys. Rev. Lett. {\bf 57,}
2879 (1986).

\bibitem{Porter}  C.E. Porter and R.G. Thomas, Phys. Rev. {\bf \ 104, }483%
{\bf \ }(1956).

\bibitem{Pato0}  M.P. Pato, C.A. Nunes, C.L. Lima, M.S. Hussein, and Y.
Alhassid, Phys. Rev. {\bf C 49, }2919{\bf \ }(1994).

\bibitem{Varga}  M. Jansen, B. Shapiro and I. Varga, Phys. Rev. Lett. {\bf %
81,} 3048 (1998).

\bibitem{Mitch}  A.A. Adams, G.E. Mitchell, and J.F. Shriner, Jr., Phys.
Lett. {\bf B 422,} 13 (1998).

\bibitem{Phmit}  M. S. Hussein, G.E. Mitchell and M.P. Pato, in preparation.
\end{references}
\end{document}